\documentstyle[psfigs]{mn}
\newcommand{\ergcms}{erg\,cm$^{-2}$\,s$^{-1}$}

\begin{document}

\title{The role of non-thermal electrons in the hydrogen and calcium lines
  of stellar flares}

\author[M.~D.~Ding and C.~Fang]
       {M.~D.~Ding and C.~Fang \\
        Department of Astronomy, Nanjing University, Nanjing 210093, China}

\date{Accepted 2001 ?????. Received 2001 ?????;
      in original form 2001 ?????}

\pagerange{\pageref{firstpage}--\pageref{lastpage}}
\pubyear{2001}

\maketitle

\label{firstpage}

\begin{abstract}

There is observational evidence showing that stellar and solar flares
occur in a similar circumstance, although the former are usually much more
energetic. It is expected that the bombardment by high energy electrons is
one of the chief heating processes of the flaring atmosphere. In this
paper we study how a precipitating electron beam can influence the line
profiles of Ly$\alpha$, H$\alpha$, Ca\,{\sc ii} K and $\lambda 8542$.
We use a model atmosphere of a dMe star and make non-LTE computations
taking into account the non-thermal collisional rates due to the
electron beam. The results show that the four lines can be enhanced
to different extents. The relative enhancement increases with increasing
formation height of the lines. Varying the energy flux of the electron
beam has different effects on the four lines. The wings of Ly$\alpha$ and
H$\alpha$ become increasingly broad with the beam flux; change of the
Ca\,{\sc ii} K and $\lambda 8542$ lines, however, is most significant in
the line centre. Varying the electron energy (i.e., the low-energy
cut-off for a power law beam) has a great influence on the Ly$\alpha$
line, but little on the H$\alpha$ and Ca\,{\sc ii} lines. An electron
beam of higher energy precipitates deeper, thus producing less
enhancement of the Ly$\alpha$ line. The Ly$\alpha$/H$\alpha$ flux ratio
is thus sensitive to the electron energy.

\end{abstract}


\section{Introduction}

Stellar flares have been frequently detected in both broad band photometry
and in spectral lines. A large stellar flare can be several orders of
magnitude more energetic than its solar analogy. During a stellar flare,
the continuum emission in UV and visible bands is significantly enhanced
and is usually characterised by a blue colour (e.g. Pagano et al. 1997);
on the other hand, spectral lines show an increased net emission, with a
magnitude varying from line to line (e.g. Catalano \& Frasca 1994; Montes
et al. 1996); lines show asymmetries and macroscopic velocity fields of
several hundreds of km s$^{-1}$ possibly exist (e.g. Houdebine et al.
1993; Gunn et al. 1994; Montes et al. 1999), implying that the flaring
atmosphere is very dynamic.

A detailed comparison between stellar and solar flares was made by Haisch,
Strong \& Rodon\`o (1991). Despite the big difference in magnitude and
scale, stellar and solar flares still show many similar features in the
enhancement of lines and continua, implying further that the basic heating
processes and emission mechanisms are similar in these two phenomena. It
is widely accepted that a solar flare occurs through the reconnection of
magnetic fields, which releases a large amount of thermal energy and
results in the acceleration of charged particles. Accelerated particles
then precipitate downwards to heat the lower atmosphere, leading to a
dynamic evolution of the atmosphere (chromospheric evaporation and
condensation). Although there is no direct information on spatially
resolved magnetic activity on stars like that observed on the Sun, some
observations still show evidence that a stellar flare may result from the
interaction between an old magnetic structure and a new emerging flux
(Catalano \& Frasca 1994). Simon, Linsky \& Schiffer (1980) proposed a
speculative scenario of major long-lived RS CVn flares in which the
component stars have very large corotating flux tubes, which occasionally
interact. Recently, Rubenstein
\& Schaefer (2000) proposed that superflares, which are newly detected on
F--G main-sequence stars by Schaefer, King \& Deliyannis (2000), are
caused by magnetic reconnection between fields of the primary star and a
close-in planet. In these cases, the basic flare mechanism is also 
magnetic reconnection.

Therefore, we can postulate that the eruption of a stellar flare involves
the acceleration of charged particles which then bombard the lower
atmosphere. This process not only provides direct heating through Coulomb
collisions, producing the bulk mass motions as revealed in the observed
spectra (e.g. Houdebine et al. 1993), but also yields non-thermal
collisional excitation and ionization of the ambient atoms. The latter
increases the source function and thus enhances the line and continuum
emission. This issue is important in the spectroscopy and, in particular,
the atmospheric modelling of stellar flares. It is then desirable to have
a quantitative assessment of the non-thermal effects on both the line and
continuum emission. Numerical computations including the non-thermal
effects have been done mostly for solar flares (e.g. Fang, H\'enoux \& Gan
1993; H\'enoux, Fang \& Gan 1995; Ding \& Fang 1997; Ding \& Schleicher
1997; Ding 1999; Fang, H\'enoux \& Ding 2000). Recently, Ding \& Fang
(2000) studied the role of non-thermal electrons in the optical continuum
of stellar flares. They found that an electron beam with a large energy
flux can produce a $U$-band brightening and a $U-B$ colour that are
roughly comparable with the observed values of a typical large flare.
This work investigates how the non-thermal electrons influence the
lines of hydrogen (H\,{\sc i} Ly$\alpha$ and H$\alpha$) and the lines of
ionized calcium (Ca\,{\sc ii} K and $\lambda 8542$) which are frequently
used in the spectroscopy of stellar atmospheres (e.g. Hawley \& Pettersen
1991; Panagi, Byrne \& Houdebine 1991; Houdebine \& Doyle 1994a, 1994b;
Houdebine \& Stempels 1997).

\section{Computational Method}

When an electron beam precipitates downwards from the corona, the beam
electrons collide elastically with the ambient electrons and protons,
leading to a direct heating of the plasma; meanwhile, the beam electrons
collide inelastically with the ambient atoms and produce a non-thermal
excitation and ionization of the atoms. As in Ding \& Fang (2000), we
first evaluate the rate of energy deposition corresponding to the latter
process, which reads
\begin{equation}
\Phi = \frac{1}{2} (1-\xi)n_{\mathrm H}\Lambda' K(\delta -2)
  \frac{{\cal F}_{1}}{E^{2}_{1}} \left( \frac{N}{N_{1}} \right)
  ^{-\frac{\delta}{2}} \int^{u_{1}}_{0}
  \frac{u^{\frac{\delta}{2}-1}{\mathrm d}u}
  {(1-u)^{\frac{2+\beta}{4+\beta}}},
\end{equation}
where $\xi$ is the ionization degree of the ambient plasma,
$K=2\pi e^{4}$, $\Lambda'$ the Coulomb logarithm for inelastic collisions.
We assume a power law flux spectrum for the electron beam, characterised
by a power index, $\delta$, a low-energy cut-off, $E_{1}$, and a total
energy flux, ${\cal F}_{1}$. $N$ is the column density and $N_{1}$ is the
depth penetrated by electrons of an energy $E_{1}$, which is expressed as
\begin{equation}
N_{1}=\frac{\mu_{0}E^{2}_{1}}{\left(2+\beta/2\right) \gamma K} ,
\end{equation}
where $\mu_{0}$ is the cosine of the initial pitch angle, taken to be
unity here. The expression for $\beta$ and $\gamma$ can be found in
Emslie (1978). Moreover, in equation (1),
\begin{equation}
u_{1}=\left\{ \begin{array}{ll}
                1,        & N>N_{1},\\
                N/N_{1},  & N<N_{1}.
              \end{array}
      \right.
\end{equation}

We adopt an atomic model of four bound levels plus continuum for hydrogen
and a model of five bound levels plus continuum for ionized calcium.  The
non-thermal excitation and ionization rates are computed as
\begin{equation}
  C^{\mathrm B}_{ij}\simeq a_{ij}\Phi/n_{1},
\end{equation}
where $n_{1}$ is the hydrogen ground level population. The coefficients
$a_{ij}$ are taken from Fang et al. (1993): in cgs units,
$a_{12}=2.94\times 10^{10}$, $a_{13}=5.35\times 10^{9}$,
$a_{14}=1.91\times 10^{9}$ and $a_{1c}=1.73\times 10^{10}$ for H;
$a_{14}=2.38\times 10^{10}$, $a_{15}=4.25\times 10^{10}$
and $a_{1c}=4.69\times 10^{10}$ for Ca\,{\sc ii}.
The above formulae for non-thermal collisional rates are derived
empirically using the collisional cross-sections for various
transitions (Fang et al. 1993). It is very convenient to include these
rates into the computation of model atmospheres.

We do non-LTE computations using a method similar to that described
in Gan \& Fang (1987) and Fang et al. (1993). The general procedure is
as follows. Given a model atmosphere and a prescribed electron beam,
we solve iteratively the equations of hydrostatic equilibrium, particle
conservation, radiative transfer and statistical equilibrium, including
the non-thermal collisional rates ($C^{\mathrm B}_{ij}$), for both H and
Ca\,{\sc ii}. After the computations reach convergence, we then calculate
the line source function and the opacity, and finally the line profile as
\begin{equation}
F_{\lambda}=2\pi\int^{1}_{0} I_{\lambda}(\mu) \mu {\mathrm d}\mu =
  2\pi\int^{1}_{0}{\mathrm d}\mu \int^{\infty}_{0}
  S_{\lambda}{\mathrm e}^{-\tau_{\lambda}/\mu}{\mathrm d}\tau_{\lambda}.
\end{equation}
Broadening mechanisms include radiative damping, the Doppler effect and
the Stark effect. Note that for all the lines involved, we assume
complete frequency redistribution (CRD), while the effect of partial
frequency redistribution (PRD) may be important for some lines
(e.g. Ly$\alpha$ and Ca\,{\sc ii} K). However, CRD is still a reasonable
assumption in the present investigation since our main purpose is to
show the effect of non-thermal electron beams, not to reproduce exactly
the observed line profiles.

\section{Influence of Electron Beams on the Line Emission}

Many authors have investigated how the spectral lines, including the
H\,{\sc i} Ly$\alpha$ and H$\alpha$, and the Ca\,{\sc ii} K and $\lambda
8542$ lines, change with the modifications of atmospheric models of, for
example, late-type stars (e.g. Linsky et al. 1979; Houdebine \&
Doyle 1994a, 1994b; Houdebine, Doyle, \& Ko\'scielecki 1995;
Houdebine \& Stempels 1997; Short \& Doyle 1997, 1998; Mauas 2000).
These lines are shown to be good
diagnostics of the atmospheric conditions in the late-type stars.
Generally speaking, the Ca\,{\sc ii} K line, in particular, the K1
minimum, can be used to determine the temperature structure around the
temperature minimum region; emission in the hydrogen Balmer lines
(H$\alpha$ etc.) is related to the temperature and temperature gradient
in the chromosphere; the Ly$\alpha$ flux relies sensitively on the
position of the transition region, i.e., the coronal pressure. However,
we will show below that the presence of a non-thermal electron beam in
a flare star can also change these spectral features without modifying
the atmospheric model. Therefore, in order to construct reliable
semi-empirical models or make proper spectral diagnostics for these flare
stars, we need first to study quantitatively the effect of
non-thermal electrons on the above-mentioned lines.

\subsection{Effect of varying the beam flux}

\begin{figure}
\centerline{\psfig{file=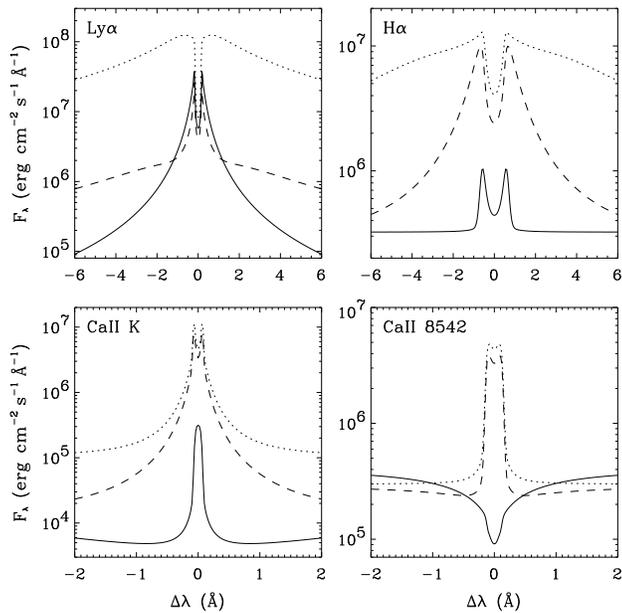, angle=0, width=8.2cm}}
\caption[]{Line profiles of Ly$\alpha$, H$\alpha$, Ca\,{\sc ii} K and
$\lambda 8542$ computed from a model atmosphere of the dMe star AD Leo
(Mauas \& Falchi 1994), considering the effect of a precipitating
non-thermal electron beam. Different line styles show the effect of
varying the energy flux, ${\cal F}_{1}$, of the beam:
0 (solid line), 10$^{10}$ (dashed line) and 10$^{12}$ (dotted line)
\ergcms. In all cases, the low-energy cut-off, $E_{1}$, is 20 keV, and
the power index, $\delta$, is 3.}
\label{fig1}
\end{figure}

As in Ding \& Fang (2000), we adopt the semi-empirical model of the dMe
star, AD Leo, proposed by Mauas \& Falchi (1994), as the base model,
and then introduce an electron beam into the atmosphere to compute the
emergent line profiles. We fix the model atmosphere by only changing the
energy flux of the beam, covering a range of 10$^{9}$--10$^{12}$ \ergcms,
which are typical values for a solar flare case. Fig.~\ref{fig1}
displays the four line profiles in different cases of beam fluxes.
Generally speaking, the four lines undergo a drastic change when the
beam flux increases. However, there is clearly a difference in the
evolutional behaviour of these lines.
When the beam is not so strong (${\cal F}_{1}\la 10^{10}$ \ergcms), the
Ly$\alpha$ line shows only an increased emission in its far wings, while
the inner part, including the emission peaks and the line centre, is not
increased but somehow decreased. When the beam flux further
increases, the whole line is enhanced significantly, except for the
line centre, which remains as weak as in the quiescent state. The
H$\alpha$ line shows instead a monotonic increase in the whole profile
with the beam flux. In the case of a strong beam (${\cal F}_{1}\sim
10^{12}$ \ergcms), both the H$\alpha$ and Ly$\alpha$ lines become very
broad. Quite differently, change of the Ca\,{\sc ii} K and $\lambda 8542$
lines is most significant in the line centre. While the wings of the K
line can also be enhanced to some extent, the wings of the $\lambda 8542$
line are slightly reduced by the non-thermal effects.

\begin{figure}
\centerline{\psfig{file=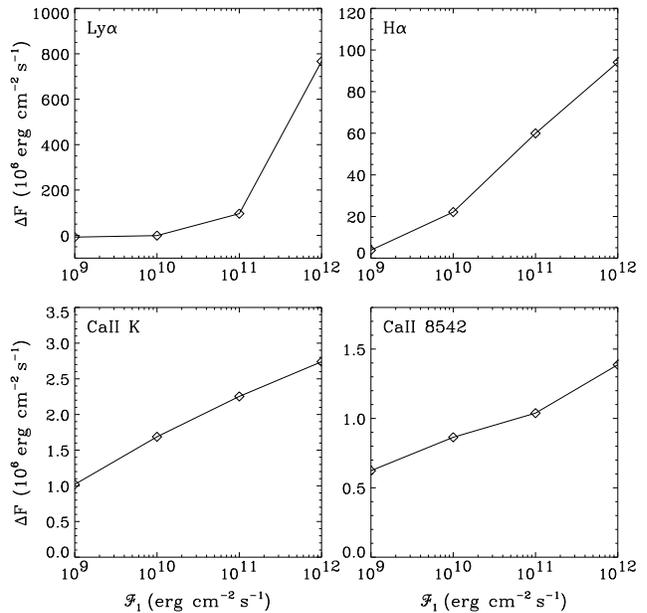, angle=0, width=8.2cm}}
\caption[]{Net integrated fluxes of Ly$\alpha$, H$\alpha$, Ca\,{\sc ii}
K and $\lambda 8542$ against the energy flux of the electron beam. In all
cases, the low-energy cut-off, $E_{1}$, is 20 keV, and the power index,
$\delta$, is 3.}
\label{fig2}
\end{figure}

To show more quantitatively the variation of the lines, we plot in
Fig.~\ref{fig2} the net integrated flux, $\Delta F$, defined as the
flux integrated over the whole line profile with the quiescent value
subtracted, against the electron beam flux. The integration extends to
$\Delta\lambda=\pm 6$ \AA\ for Ly$\alpha$ and H$\alpha$ lines and to
$\Delta\lambda=\pm 2$ \AA\ for Ca\,{\sc ii} K and $\lambda 8542$ lines,
respectively. It can be seen
that the extent of line flux enhancement increases in the order of
Ca\,{\sc ii} $\lambda 8542$, K, H$\alpha$ and Ly$\alpha$. From line
profile computations, we can find that this is just the order of line
formation heights, spanning a range from the upper photosphere or lower
chromosphere to the upper chromosphere. So the results are compatible
with the general picture that the electron beam precipitates downwards
from the corona. We note that the negative value of $\Delta F$ for
Ly$\alpha$ in weak beam cases is due to a reduction of emission in the
inner part of the line, as described above.

\subsection{Effect of varying the electron energy}

\begin{figure}
\centerline{\psfig{file=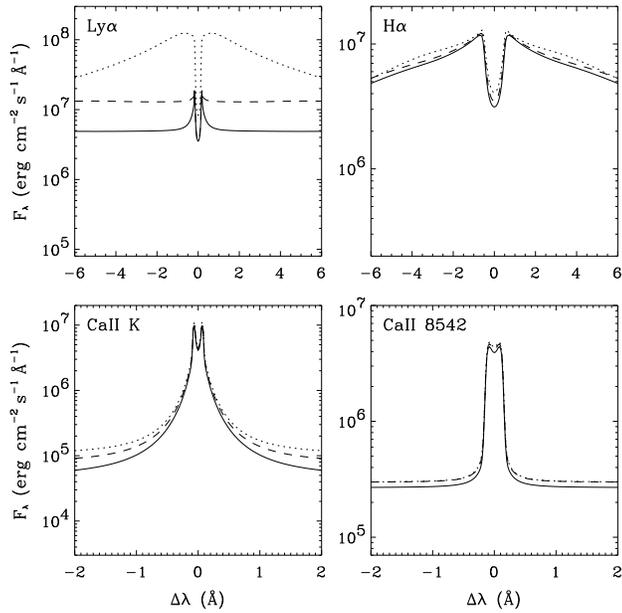, angle=0, width=8.2cm}}
\caption[]{Line profiles of Ly$\alpha$, H$\alpha$, Ca\,{\sc ii} K and
$\lambda 8542$ computed from a model atmosphere of the dMe star AD Leo
(Mauas \& Falchi 1994), considering the effect of a precipitating
non-thermal electron beam. Different line styles show the effect of
varying the low-energy cut-off, $E_{1}$, of the beam: 20 (dotted
line), 60 (dashed line) and 100 (solid line) keV. In all cases, the
energy flux, ${\cal F}_{1}$, is 10$^{12}$ \ergcms, and the power index,
$\delta$, is 3.}
\label{fig3}
\end{figure}

\begin{figure}
\centerline{\psfig{file=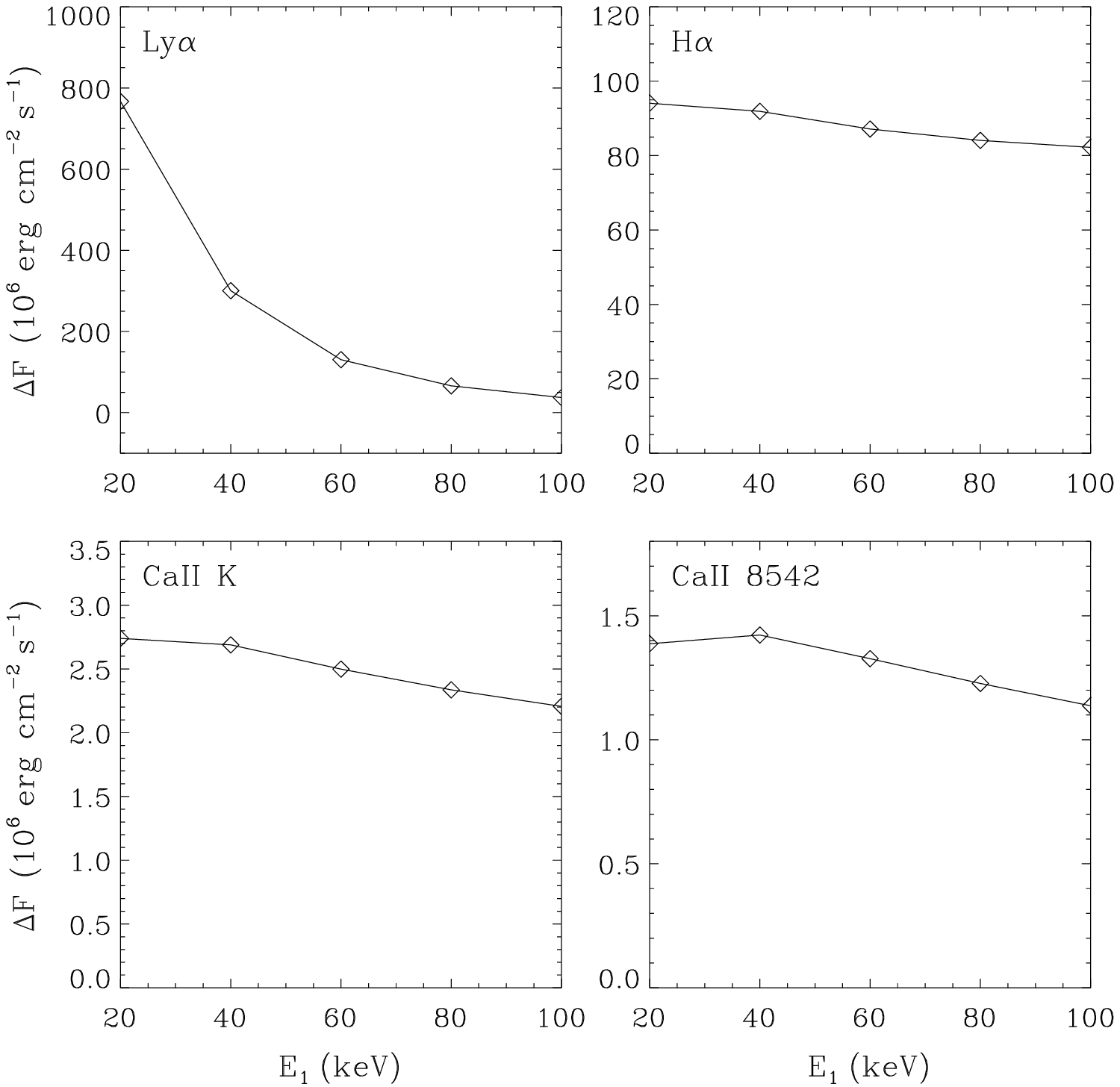, angle=0, width=8.2cm}}
\caption[]{Net integrated fluxes of Ly$\alpha$, H$\alpha$, Ca\,{\sc ii}
K and $\lambda 8542$ against the low-energy cut-off of the electron beam.
In all cases, the energy flux, ${\cal F}_{1}$, is 10$^{12}$ \ergcms, and
the power index, $\delta$, is 3.}
\label{fig4}
\end{figure}

To check the effect of non-thermal electrons of different energies, 
we have also computed the line spectra for electron beams with a fixed
energy flux but different low-energy cut-offs. The line profiles and
the net integrated fluxes are plotted in Figs.~\ref{fig3} and \ref{fig4}
respectively. Fig.~\ref{fig3} shows that, varying $E_{1}$ does not alter
obviously the profiles
of H$\alpha$, Ca\,{\sc ii} K and $\lambda 8542$. However, it has a great
influence on the Ly$\alpha$ profile, namely, increasing $E_{1}$ reduces
the non-thermal effects on the line intensities. This result is not
surprising. As the Ly$\alpha$ line is formed in the very upper
chromosphere, electrons of a higher energy penetrate deeper and thus
leave less energy in upper layers to excite the hydrogen atoms.

In the above computations, the electron beam is assumed to originate in
the corona, i.e., at the top of the model atmosphere. If the beam
originates in a lower layer, its effect could become more significant
for lines formed in the lower atmosphere. Such a scenario has been
proposed to be a possible origin of a flare-like phenomenon on the Sun,
Ellerman bombs (Ding, H\'enoux \& Fang 1998).

\subsection{Discussion}

\begin{figure}
\centerline{\psfig{file=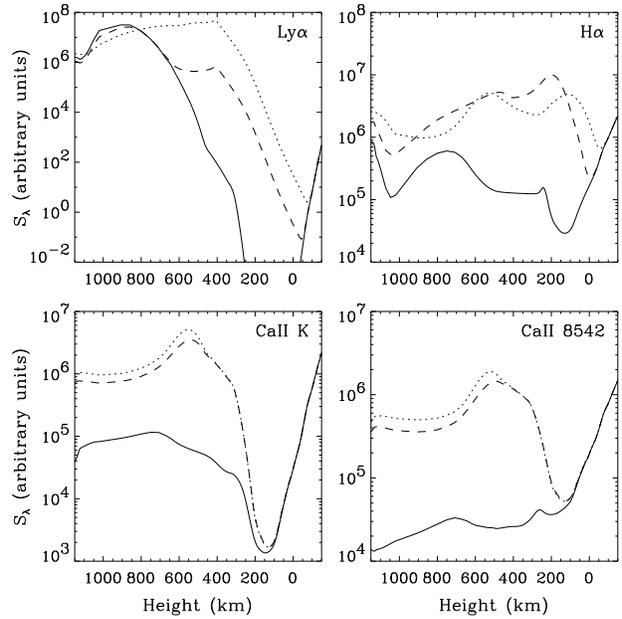, angle=0, width=8.2cm}}
\caption[]{Height distribution of the source function of the four lines
in the atmosphere of AD Leo in the presence of an electron beam. The
line styles have the same meaning as in Fig.~\ref{fig1}.}
\label{fig5}
\end{figure}

\begin{figure}
\centerline{\psfig{file=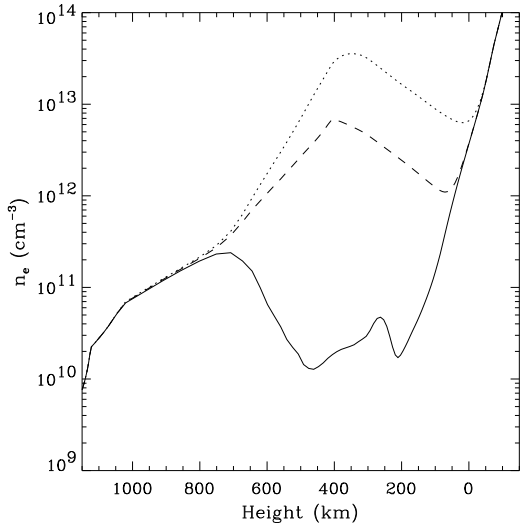, angle=0, width=7.0cm}}
\caption[]{Height distribution of the electron density in the atmosphere
of AD Leo in the presence of an electron beam. The line styles have the
same meaning as in Fig.~\ref{fig1}.}
\label{fig6}
\end{figure}

To illustrate the different behaviours of the four lines in response
to the electron beam, we plot in Fig.~\ref{fig5} the height distribution
of the line source function in different cases of beam fluxes.
It shows that the beam can make the source function of Ly$\alpha$
greatly increased in a broad region from the chromosphere to the
photosphere. This can be understood that the hydrogen ground level is
depopulated while the excited levels are overpopulated with respect to
the beam-free case. Accordingly, the Ly$\alpha$ wings are significantly
enhanced. In the very upper layers, however, the source function of
Ly$\alpha$ could be slightly reduced, which results in a decrease of
the line centre intensity in some cases (see Fig.~\ref{fig1}). Similar
results have been obtained for the Ly$\alpha$ line in solar flares
(H\'enoux et al. 1995). The source function of H$\alpha$ is also
increased but less significantly than Ly$\alpha$. Note that the Stark
effect, due to an enhanced electron density, plays an important role
in the broadening of the H$\alpha$ line. Fig.~\ref{fig6} displays the
height distribution of the electron density, which is shown to be raised
by up to three orders of magnitude in the chromosphere, highlighting the
non-thermal ionization effect of the beam.

The increase of the source function of hydrogen lines in photospheric
layers is mainly caused by a backwarming effect, i.e., the enhanced
radiation from the chromosphere, while the direct penetration of
non-thermal electrons has less effect.
In contrast, the source function of the Ca\,{\sc ii} lines is mainly
increased in the chromosphere; it is still coupled to the local Planck
function in the photosphere. In addition, the Ca\,{\sc ii} lines are
less affected by the Stark effect. Therefore, the enhancement of these
lines is most apparent in the line centre, as shown in Fig.~\ref{fig1}.

The quantitative results of line and continuum enhancement due
to an electron beam are certainly model-dependent.
In all the above computations, we have assumed a cool background
atmosphere. In fact, the model atmosphere should change with the
activity level of the star, especially when flare occurs. A model of
the flaring state of AD Leo, for example, was proposed by Mauas \&
Falchi (1996). Compared with the model in the quiescent state, a flare
model has a higher temperature in the chromosphere and even in the
photosphere, and a lower position of the transition region (a higher
coronal pressure). This model itself can produce an enhanced line
emission. However, a higher coronal mass in the flaring atmosphere
would consume a part of the electron energy, so that the effect of
the electron beam is less significant than in the case of a quiescent
atmosphere. The real situation might be a combination of the effects
of a heated atmosphere and a non-thermal electron beam.

In addition to the line flux, we find that the Ly$\alpha$/H$\alpha$
flux ratio is in positive dependence on the electron beam flux, but,
apparently, in negative dependence on the low-energy cut-off. In the case
of ${\cal F}_{1}=10^{12}$ \ergcms, $E_{1}=20$ keV and $\delta =3$, the
ratio amounts to $\sim 6$. Therefore, it seems that an electron beam can
produce most spectral features that are usually ascribed to a modification
of the model atmosphere. As long as the electron beam exists during the
flaring process, its effect should be taken into account in order to
properly construct the semi-empirical models, otherwise the chromospheric
temperature will be overestimated.

\section{Conclusions}

There is observational evidence showing that a stellar flare may occur in
a similar circumstance to the eruption of a solar flare. A common flare
scenario is the reconnection of magnetic
fields. Although a stellar flare, in particular, the flares in late-type
stars, can be several orders of magnitude more energetic than a solar
flare, their emission features, in both lines and continua, still share
many similarities. It can be postulated that the flaring process in a
star also involves the acceleration of energetic particles (protons or
electrons), carrying most energy to heat the lower atmosphere.

We study in this paper how a precipitating electron beam can influence the
line profiles of Ly$\alpha$, H$\alpha$, Ca\,{\sc ii} K and $\lambda 8542$.
We make non-LTE computations taking into account the non-thermal
collisional rates due to the electron beam. The results show that the
four lines can be enhanced to different extents. The relative enhancement
increases with increasing formation height of the lines. Varying the
energy flux of the electron beam has different effects on the four lines.
The wings of Ly$\alpha$ and H$\alpha$ become increasingly broad with the
beam flux; change of the Ca\,{\sc ii} K and $\lambda 8542$ lines, however,
is most significant in the line centre. Varying the electron energy
(i.e., the low-energy cut-off for a power law beam) has a great influence
on the Ly$\alpha$ line, but little on the H$\alpha$ and Ca\,{\sc ii}
lines. An electron beam of higher energy precipitates deeper, thus
producing less enhancement of the Ly$\alpha$ line. The
Ly$\alpha$/H$\alpha$ flux ratio is thus sensitive to the electron energy.
Together with our computations for the continuum emission (Ding \& Fang
2000), these results provide a diagnostic tool for the non-thermal
processes in stellar flares.

\section*{acknowledgements}
We would like to thank the referee for his/her valuable comments on
the manuscript. This work was supported by TRAPOYT, National Natural
Science Foundation of China under grant 10025315 and National Basic
Research Priorities Programme under grant G2000078402.

\label{lastpage}


\begin{thebibliography}{}

\bibitem{} Catalano S., Frasca A., 1994, A\&A, 287, 575
\bibitem{} Ding M. D., 1999, A\&A, 351, 368
\bibitem{} Ding M. D., Fang C., 1997, A\&A, 318, L17
\bibitem{} Ding M. D., Fang C., 2000, MNRAS, 317, 867
\bibitem{} Ding M. D., H\'enoux J.-C., Fang C., 1998, A\&A, 332, 761
\bibitem{} Ding M. D., Schleicher H., 1997, A\&A, 322, 674
\bibitem{} Emslie A. G., 1978, ApJ, 224, 241
\bibitem{} Fang C., H\'enoux J.-C., Ding M. D., 2000, A\&A, 360, 702
\bibitem{} Fang C., H\'enoux J.-C., Gan W. Q., 1993, A\&A, 274, 917
\bibitem{} Gan W. Q., Fang C., 1987, Chinese Astron. Astrophys., 11, 49
\bibitem{} Gunn A. G., Doyle J. G., Mathioudakis M., Houdebine E. R.,
  Avgoloupis S., 1994, A\&A, 285, 489
\bibitem{} Haisch B., Strong K. T., Rodon\`o M., 1991, ARA\&A, 29, 275
\bibitem{} Hawley S. L., Pettersen B. R., 1991, ApJ, 378, 725
\bibitem{} H\'enoux J.-C., Fang C., Gan W. Q., 1995, A\&A, 297, 574
\bibitem{} Houdebine E. R., Doyle J. G., 1994a, A\&A, 289, 169
\bibitem{} Houdebine E. R., Doyle J. G., 1994b, A\&A, 289, 185
\bibitem{} Houdebine E. R., Doyle J. G., Ko\'scielecki M., 1995, A\&A,
  294, 773
\bibitem{} Houdebine E. R., Foing B. H., Doyle J. G., Rodon\`o M., 1993,
  A\&A, 274, 245
\bibitem{} Houdebine E. R., Stempels H. C., 1997, A\&A, 326, 1143
\bibitem{} Linsky J. L., Hunten D. M., Sowell R., Glackin D. L.,
  Kelch W. L., 1979, ApJS, 41, 481
\bibitem{} Mauas P. J. D., 2000, ApJ, 539, 858
\bibitem{} Mauas P. J. D., Falchi A., 1994, A\&A, 281, 129
\bibitem{} Mauas P. J. D., Falchi A., 1996, A\&A, 310, 245
\bibitem{} Montes D., Saar S. H., Collier Cameron A., Unruh Y. C.,
  1999, MNRAS, 305, 45
\bibitem{} Montes D., Sanz-Forcada J., Fern\'andez-Figueroa M. J.,
  Lorente R., 1996, A\&A, 310, L29
\bibitem{} Pagano I., Ventura R., Rodon\`o M., Peres G., Micela G.,
  1997, A\&A, 318, 467
\bibitem{} Panagi P. M., Byrne P. B., Houdebine E. R., 1991, A\&AS,
  90, 437
\bibitem{} Rubenstein E. P., Schaefer B. E., 2000, ApJ, 529, 1031
\bibitem{} Schaefer B. E., King J. R., Deliyannis C. P., 2000, ApJ,
  529, 1026
\bibitem{} Short C. I., Doyle J. G., 1997, A\&A, 326, 287
\bibitem{} Short C. I., Doyle J. G., 1998, A\&A, 336, 613
\bibitem{} Simon T., Linsky J. L., Schiffer F. H., III, 1980, ApJ, 239, 911

\end{thebibliography}
\end{document}